\documentclass[final,pre,twocolumn,superscriptaddress,floatfix]{revtex4-1}

\usepackage{amsmath,amssymb,color}
\usepackage{graphicx} 
\usepackage{color}

\usepackage{siunitx}
\usepackage[]{units}

\newcommand{\abs}[1]{\left\lvert#1\right\rvert}

\begin{document}

\title{A model of chromosomal loci dynamics in bacteria as
fractional diffusion with intermittent transport.  }

\author{Marco Gherardi}
\email{marco.cosentino-lagomarsino@upmc.fr} \affiliation{Sorbonne
  Universit\'es, UPMC Univ Paris 06, Paris, France}
\affiliation{Physics Department, University of Milan, Via Celoria
  16, Milano, Italy}
\affiliation{I.N.F.N., Milano, Italy}

\author{Ludovico Calabrese}
\affiliation{Physics Department, University of Milan, Via Celoria
  16, Milano, Italy}

\author{Mikhail Tamm} 
\affiliation{Physics Department, University of Moscow, Moscow, Russia}

\author{Marco {Cosentino Lagomarsino}}
\email{marco.cosentino-lagomarsino@upmc.fr} \affiliation{Sorbonne
  Universit\'es, UPMC Univ Paris 06, Paris, France}
\affiliation{CNRS, UMR 7238, Paris, France}
\affiliation{IFOM, FIRC Institute of Molecular Oncology, Milan, Italy}


\begin{abstract}
  The short-time dynamics of bacterial chromosomal loci is a mixture
  of subdiffusive and active motion, in the form of rapid relocations
  with near-ballistic dynamics.
  While previous work has shown that such rapid motions are
  ubiquitous, we still have little grasp on their physical nature, and
  no positive model is available that describes them.
  Here, we propose a minimal theoretical model for loci movements as a
  fractional Brownian motion subject to a constant but intermittent
  driving force, and compare simulations and analytical calculations
  to data from high-resolution dynamic tracking in \emph{E.~coli}.
  This analysis yields the characteristic time scales for
  intermittency.  Finally, we discuss the possible shortcomings of
  this model, and
  show that an increase in the effective local noise felt by the chromosome 
  associates to the active relocations.
\end{abstract}

\maketitle

\section{Introduction}

%

The motion of chromosomal loci of the bacterium \emph{Escherichia
  coli} on time scales 0.1--100s shows intriguingly complex
patterns~\cite{Kleckner2014}.  These fluctuations contain key evidence
on the complex physical nature of the intracellular crowded medium
made of genome and cytoplasm~\cite{Lagomarsino2015,benza2012}, a
stimulating riddle of soft-matter physics with large biological
significance.  On top of a basal subdiffusive
dynamics~\cite{Javer2013,theriot10,theriot10b}, the loci are subject
to active forces that have been characterized as both active
noise~\cite{Weber2012} and rapid excursions of near-ballistic
nature~\cite{Javer2014,Fisher2013,joshi11}. The nature of the
background motion and of the active relocation is still under debate,
and both issues are likely deeply connected with the recent finding
that the bacterial cytoplasm shows some glass-like
properties~\cite{Parry2014}.

The rapid relocations emerge as distinct from the subdiffusive
background motion of the loci.  It is widely accepted that the
background motion is compatible with fractional Brownian motion (fBm)
or fractional Langevin behavior, i.e. with directionally
anti-correlated steps that witness viscoelastic behavior, and with a
mobility that is locus- and
cell-cycle-dependent~\cite{Javer2013,theriot10,Weber2012}.
Consequently, a precise quantification of the fact that rapid
relocations deviate from the expected behavior of pure viscoelastic
subdiffusion is possible by comparing the behavior of experimental
single tracks to a parameter-matched fractional Brownian
motion~\cite{Javer2014}. However, no positive model describing such
rapid relocations is currently available. Building such a description
is important to address relevant outstanding questions on the nature
of the driving force and the relevant time scales that play a role in
the process.  To this aim, a physical model may be difficult at
  this stage. The main obstacles are that we still know very little
  about both the nature of cytoplasmic
  diffusion~\cite{Parry2014,Lampo2017} and the nature of the
  nonequilibrium forces driving the
  chromosome~\cite{theriot12,Javer2014}. In addition, the contribution
  of chromosome folding to subdiffusion is an open
  question~\cite{theriot10b,Lampo2016,Polovnikov2017}. In this
  context, a phenomenological model realizing the main pertinent
  features can be an useful first step~\cite{Espeli2017}.

Importantly, the trajectories showing rapid movements in these data
clearly produce superlinear behavior of the mean-square displacement
(MSD)~\cite{Javer2014}.
It is well known that in the case of viscoelastic subdiffusion, an
object under constant driving force has to produce a sublinear
mean-square displacement (drift), in order to follow a
fluctuation-dissipation
relation~\cite{Lutz:2001,Barkai1998,Espeli2017}. This constraint is
realized by the fractional Langevin
equation~\cite{Lutz:2001,Taloni2010,Kuwada2013}. Thus, given the
superdiffusive stretches of motion over a sub-Rouse basal diffusion,
it is not clear whether a Stokes-Einstein relation should apply in
this situation.
The validity of a generalized Einstein relation is ultimately due to
the precise nature of the drive, which in this case may act both on
the probe and on the environment. For example, a possibility is that
the rapid relocations are due to large-scale rearrangements of the
chromosome~\cite{Fisher2013,Javer2014}.

The recent theoretical literature has focused on active noise,
modeled as colored fluctuations violating the fluctuation-dissipation
theorem, and on its effect on a Rouse model in a viscoelastic
medium~\cite{Sakaue2016,Vandebroek2015}.
However, this framework cannot capture the ballistic stretches
observed in the \emph{E.~coli} data. In this system, rapid relocations
emerge as distinct from the subdiffusive background motion of the
loci, as can be seen by comparing experimental single tracks to a
parameter-matched fractional Brownian motion (fBm)~\cite{Javer2014}.

Here, we take the complementary assumption
that active behavior is due to  an intermittent force.
For simplicity, we give up the description of the polymer degrees of
freedom and concentrate on the superposition of subdiffusion with an
intermittent driving force. This approach cannot explicitly
  address the stress propagation between different
  chromosomal loci, measurable from joint tracking, which is being
  addressed in the current literature for the equilibrium
  case~\cite{Polovnikov2017,Lampo2016}.
Other modeling approaches in the recent literature have explicitly
  described the dynamics of an active polymer, but represented the
  active drive as nonthermal noise or as contact of different sets of
  monomers with two different
  thermostats~\cite{Vandebroek2015,Grosberg2015,Osmanovic2017}.
%
%
%
We define a minimal description of the movement of chromosomal loci as
a basal fractional Brownian motion superposed to an intermittent
process imposing a constant driving force. Comparing with
high-resolution tracking data in \emph{E.~coli}, we set out to
identify the key behaviors and relevant parameters.

Similar intermittent processes have been proposed in the literature as
models for search processes~\cite{Loverdo2009} and complex
reaction-diffusion~\cite{Benichou2008, Oshanin2010, Benichou2010}, as
well as for describing the interplay between diffusion and active
transport in
cells~\cite{Brangwynne2009,Lagache2008,Parmeggiani2004}. However, the
case of the superposition between subdiffusion and active transport
has never been considered, giving a wider motivation to our
investigation.

\section{Model: intermittent transport and subdiffusion.}

\begin{figure}[t]
  \centering
 \includegraphics[width=0.4\textwidth]{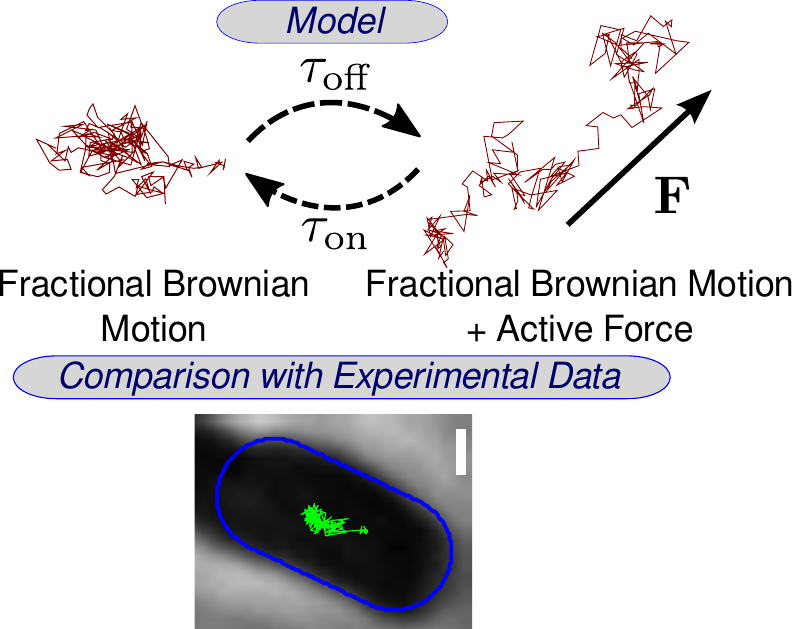}
 \caption{Illustration of the model. An intermittent process
     switches on and off a constant driving force, acting on top of a basal
     subdiffusive process with viscoelastic-like behavior, modeled as
     fractional Brownian motion. The resulting tracks are compared
     with experimental data from Ref.~\cite{Javer2014}. }
  \label{fig:1}
\end{figure}

Fig.~\ref{fig:1} illustrates the basic ingredients of the model. A
point particle is subject to fractional Gaussian
noise~\cite{Deng2009}, which captures the viscoelastic-like
subdiffusion~\cite{Weber2012} of chromosomal loci, and to an external
driving velocity of constant intensity and orientation, but active
only at intermittent intervals.

In order to reproduce near-ballistic motion in the driven stretches of
motion, this phenomenological model gives up the Einstein relation and
uses fractional Brownian motion as a model for the basal subdiffusion.
The combination of these ingredients produces the following Langevin-type 
equation of motion (for each, assumed independent, coordinate $x_i$):
\begin{equation}
  \dot{x_i}(t) = \xi_i^H(t) + V_i\sigma(t) \ ,
  \label{eq:modello}
\end{equation}
where $\mathbf{V}$ is the constant velocity caused by an external
force (drag and temperature are incorporated in $\xi_i^H$).  The
fractional noise $\xi_i^H(t)$ is a Gaussian noise with the following
correlation properties:
\begin{equation}
\label{eq:noise}
\begin{split}
  \left< \xi_i^H(t) \right> &= 0;\\
  \left< \xi_i^H(t) \xi_j^H(t') \right> &=
  2D_\mathrm{app}H(2H-1)|t-t'|^{2H-2}\delta_{ij}\\ 
  & \phantom{=}+ 4D_\mathrm{app}H |t-t'|^{2H-1}\delta(t-t')
  \delta_{ij} \ ,
\end{split}
\end{equation}
where $H$ is the Hurst exponent (or coefficient) and $D_\mathrm{app}$
is the apparent diffusion constant.  When $H<1/2$ noise at different
times is anti-correlated, with power-law relaxation (see
Ref.~\cite{Qian2003} for a justification of the expressions in
Eq.~\ref{eq:noise}).
The stochastic process $\sigma(t)$ governing how the external force
switches on and off is a standard dichotomous telegraph process, with
states 0 and 1. It is specified by the two characteristic switching
times $\tau_{\mathrm{on}}=w \tau_0$ and
$\tau_{\mathrm{off}} = (1-w) \tau_0$ bringing the system from the on
state to the off state and viceversa, respectively. In what follows,
when most convenient, we use the variables $w$ (average fraction of
time in the ``on'' state) and $\tau_0$ (average length of a full
``off''-``on''-``off'' cycle) to characterize this process.

The probability $P_+$ and $P_-$ that the force is switched on or off
at time $t$, i.e.~that $\sigma(t) = 1$ or $\sigma(t) = 0$
respectively, obeys the following master equation:
\begin{displaymath}
  \partial_t P_\pm(t) = 
  \mp \frac{P_+(t)}{w \tau_0} \pm \frac{P_-(t)}{(1-w) \tau_0} \ . 
\end{displaymath}
In the stationary state of the telegraph process the probability of
the ``on'' state equals $P_+^s = w$ (clearly, the probability of the
``off'' state is $P_-^s = 1-w$).  We assume this stationary state as
the initial condition in what follows.

\section{Results}

\subsection*{Analytical form of the mean-square displacement in
  presence of active forces}

Our first result is the derivation of an exact analytical expression
for the mean-square displacement of the model.  As we will show, this
expression is very useful to compare the model to experimental data.

The equation of motion of the process, Eq.~\eqref{eq:modello}, can be
integrated formally, obtaining
\begin{displaymath}
\Delta x_i(\tau) = \Delta x_{i,\mathrm{fBm}}(\tau)+ V_iT_{\mathrm{on}}(\tau) \ , 
\end{displaymath}
where $\Delta x_i(\tau)$ stands for the net change of a coordinate
from the initial condition, $\Delta x_{i,\mathrm{fBm}}(\tau)$ is the
contribution of the fractional noise, and $T_{\mathrm{on}}(\tau) =
\int_0^{\tau}\sigma(t)\,dt$ is a random variable representing the time
during which the active force is switched on in the interval $[0,
\tau]$.
Considering the square of this expression and averaging leads to an
expression for the mean-square displacement.  By observing that mixed
terms average to zero, due to the random force and the telegraph
process being independent, and that the mean signed step of a
fractional Brownian motion is null, we obtain
\begin{equation}
\mathrm{MSD}(\tau) = 2 d D_\mathrm{app}\tau^{2H}+ V^2 \langle
T_{\mathrm{on}}^2(\tau) \rangle \ , 
\label{eq:MSDformal}
\end{equation}
where $d$ is the dimensionality of the space.
We choose $d=2$, as the nature
  of essentially all experimental tracking data is intrinsically
  two-dimensional, although loci move in three dimensions.  

The computation of $\langle T_{\mathrm{on}}^2(\tau) \rangle$ relies on
the fact that the correlation function for a telegraph process is
known:
\begin{equation*}
\begin{split}
  \langle \sigma(t_1)\sigma(t_2) \rangle &= \langle \sigma_\mathrm{s}
  \rangle^2 + \mathrm{Var} \left( \sigma_\mathrm{s} \right) \exp
  \left(-\frac{\abs{t_2-t_1}}{\tau_\mathrm{c}}\right) \\
  &= w^2 + w(1-w) \exp
  \left(-\frac{\abs{t_2-t_1}}{\tau_\mathrm{c}}\right)\ ,
\end{split}
\end{equation*}
where we used the shorthand $\tau_\mathrm{c}=w(1-w)\tau_0$ for the
time scale in the exponential.  The second equality uses the
analytical results for the average and the variance of $\sigma$ in the
stationary state.
The mean square waiting time can then be obtained by double
integration as $\langle T_{on}^2(\tau) \rangle = \iint_\Omega\langle
\sigma(t_1)\sigma(t_2)\rangle\,\mathrm{d}t_1\mathrm{d}t_2$ within the
square $\Omega = \{0\leq t_1\leq \tau, 0\leq t_2\leq \tau \}$.
By substituting the result into Eq.~\eqref{eq:MSDformal} we obtain the
final analytical expression for the mean-square displacement in
presence of the switching active force,
\begin{equation}
\begin{split}
  \mathrm{MSD}(\tau) &= 2dD_\mathrm{app}\tau^{2H} + V^2 w^2 \tau^2
  \\
  &+ 2 V^2 \frac{\tau_0^2}{w(1-w)} 
  \left(e^{-\tau/\tau_\mathrm{c}} - 1 + \frac{\tau}{\tau_\mathrm{c}}\right).
\label{eq:parametri}
\end{split}
\end{equation}
For small lag times, namely when $\tau \ll \tau_\mathrm{c}$, the
linear term in parentheses cancels out with the term coming from the
first-order expansion of the exponential, giving
\begin{equation}
\mathrm{MSD}(\tau) \approx 
V^2 w^2 \tau^2 + 2dD_\mathrm{app}\tau^{2H} \ .
\label{eq:parametri2}
\end{equation}
The above formula shows that in the limit of small lag times,
  since the lower power $2H$ dominates, the process is increasingly
  similar to a fractional Brownian motion with Hurst exponent $H$.

\subsection*{Quantitative agreement between model and data}

\begin{figure}[t!]
  \centering
  \includegraphics[width=0.45\textwidth]{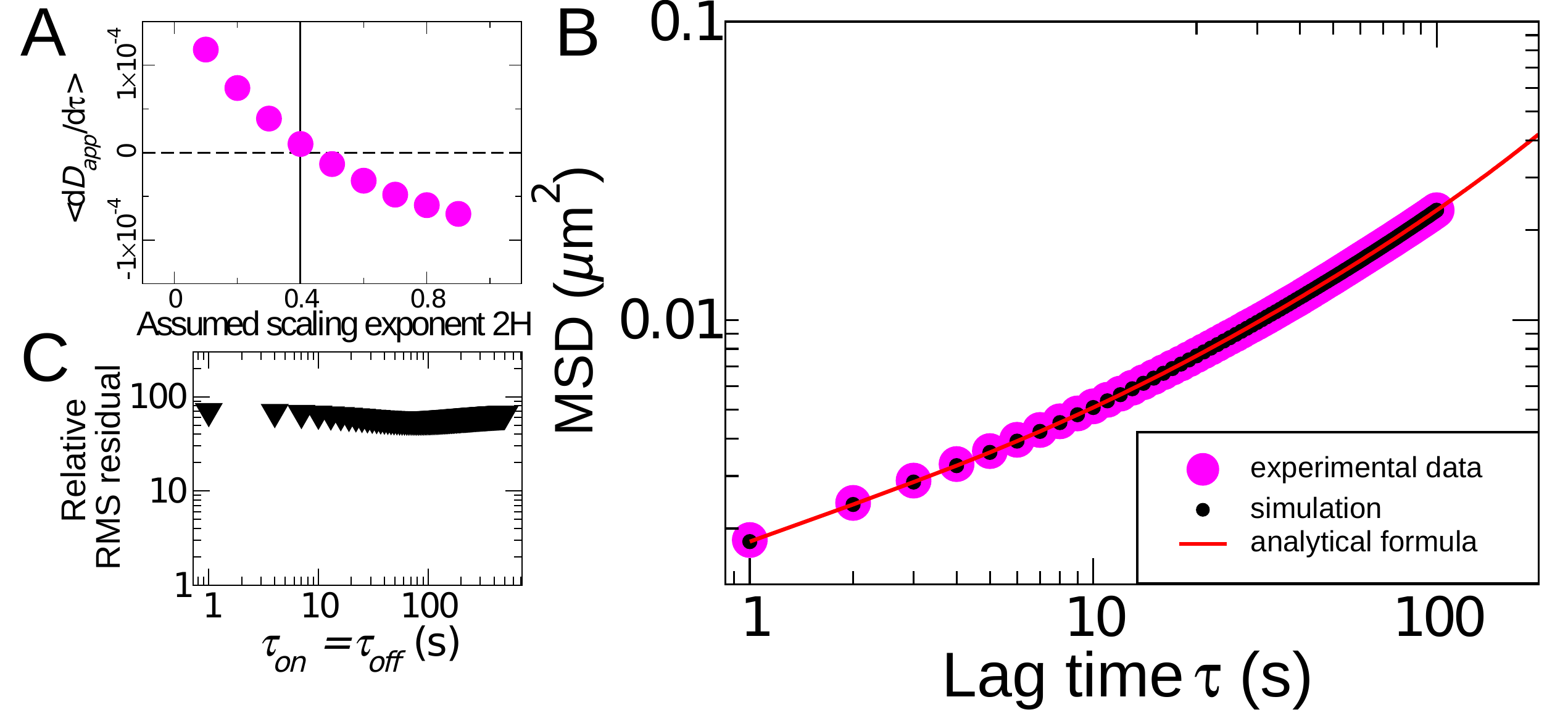}
  \caption{The model reproduces the mean-square displacement of
    experimental data allowing to fix all model
    parameters. A. Determination of the scaling exponent $2H$ of the
    background subdiffusion.  Purple circles are the mean of the
    discrete derivative (1s steps) of $D_{\mathrm{app}}:=
    \left\langle r^2(\tau)/t^{2H}\right\rangle$, for values of lag
    time below 10 seconds, where the averages are not affected by the
    influence of ballistic stretches~\cite{Javer2014}. Since a
    well-defined $D_{\mathrm{app}}$ should not vary with time scale,
    the optimal estimated value for the scaling exponent $H\simeq 0.2$
    is obtained when the plot crosses zero.  B: Large purple circles
    represent mean-square displacement vs lag time for the data on the
    Ori3 locus in Ref.~\cite{Javer2014} (error bars, SE, are smaller
    than symbols). Red line and small black circles are, respectively,
    a fit with the analytical formula, Eq.~(\ref{eq:parametri}), and
    simulation results obtained with the parameter values fixed by the
    fits (reported in table \ref{tab:params1}A).  $2H=0.41$ is fixed from
    the analysis shown in panel A. C: Fits with $w = 1/2$ cannot
    reproduce the data. The plot shows the ratio of the RMS residual
    of fits imposing $w = 1/2$ to the RMS residual of the
    unconstrained fit shown in panel B, for different values of the
    time scale. The ratio between the two values is in the range
    50--70.  }
  \label{fig:2}
\end{figure}

We used the analytical expression for the mean-square displacement to
fix model parameters from experimental data. In order to do so, we
first determined separately an optimal value of the subdiffusion
exponent $2H$ for the background properties. To obtain this value we
defined a procedure based on the fact that for time lags below $\sim
10$s, the mean-square displacements are compatible with pure
subdiffusion~\cite{Javer2013,Javer2014}. Hence, by assuming varying
values of $H$, we looked at the coherence of $4D_{\mathrm{app}}(\tau)
:= \mathrm{MSD}(\tau)/\tau^{2H}$, by taking the mean of its derivative
with respect to $\tau$ and verifying where it crossed zero
(Fig.~\ref{fig:2}A). This procedure gives a value $2H\simeq0.4$, in
line with previous studies focused on the subdiffusion of
\emph{E.~coli} chromosomal loci~\cite{theriot10,Javer2013}.
All our fits (including the ones described below) were performed
  with a nonlinear least-squares Marquardt-Levenberg algorithm
  minimizing the chi-square residual, weighted on the standard errors
  of the input data.

\begin{table}
  \centering
\[
\begin{array}{r|ccccc|cc}
\mathrm{param.} &  \boldsymbol{2H} & \boldsymbol{4D_{\textrm{app}}} & \boldsymbol{V} 
& \boldsymbol{\tau_\mathrm{on}} & \boldsymbol{\tau_\mathrm{off}}
& \boldsymbol{\tau_0} & \boldsymbol{w} \\
\mathrm{units} &  \mathrm{none} &  \unitfrac[]{{\mu m}^2}{{s}^{2H}} & \unitfrac[]{\mu m}{s}  & \mathrm{s} & \mathrm{s}
& \mathrm{s} &  \mathrm{none} \\
\hline
\mathrm{fits:}\;\;\mathrm{A} &  \approx 0.4 & 0.00182 & 0.023 & 7 & 427 & 434 & 0.016  \\
\mathrm{B} & 0.40 & 0.00180 & 0.024 & 7.0 & 443 & 450 & 0.016 \\
\mathrm{C} & 0.404 & 0.00182 & 0.020 & 7.7 & 412 & 420 & 0.018\\ 
\mathrm{D} & 0.48 & 0.00167 & 0.005 & 27 & 377 & 404 & 0.067 \\
\mathrm{E} & 0.79 & 0.00063 & \mathrm{NA} & \mathrm{NA} & \mathrm{NA} & \mathrm{NA} & \mathrm{NA} \\
\hline

\end{array}
\]
\caption{A: Parameter values for the theoretical curve and simulations
  of the constrained model fit shown in Fig.~\ref{fig:2}.  B:
  Parameter values from the alternative fitting procedure using the
  subtraction of the subdiffusing contribution to the MSD, shown in
  Fig.~\ref{fig:3}.
  C: Parameter values for the systematic 4-parameter fit.
  D: Parameter values for the joint model fit considering the
  end-to-end distance distribution. E: Parameters of the best-fitting
  fBm. (Estimated errors affect the last displayed digits.)}
\label{tab:params1}
\end{table}

Subsequently, we performed a 4-parameter fit of the data with
Eq.~(\ref{eq:parametri}), thus fixing $\tau_0$, $w$, $V$, and
$D_\mathrm{app}$.  The parameter values obtained are reported in
Table~\ref{tab:params1}A.
We also verified that direct simulations of the model with this choice
of parameters agreed with the data.  Figure \ref{fig:2}B shows that
the agreement between the mean-square displacement curves given by the
analytical formula, by simulations of the model, and by data is
excellent.  The ingredients of the model are therefore sufficient to
reproduce the experimental observations on the mean-square
displacement.

Importantly, a model with a reduced parameter space, where the two
characteristic switching times (on and off) are equal, \emph{cannot}
reproduce the mean-square displacement of the experimental data.  We
considered a constrained fit with $w=1/2$, by varying this parameter
in a wide range of values.  The results, shown in Fig~\ref{fig:2}C,
clearly indicate that the performance of this fit is much worse.
Intuitively, if $\tau_\mathrm{on}=\tau_\mathrm{off}$ the number of
tracks in which the force is off all the time (i.e., the purely
subdiffusive ones) equals the number of tracks in which it is always
switched on. 
However, it is clear that the tracks showing simple subdiffusive
behavior are much more common in the data than the tracks dominated by
ballisitic-like drift~\cite{Javer2014}.

\subsection*{Subtraction of subdiffusive noise unveils superdiffusive
  behavior in the data}

\begin{figure}[t!]
  \centering
  \includegraphics[width=0.35\textwidth]{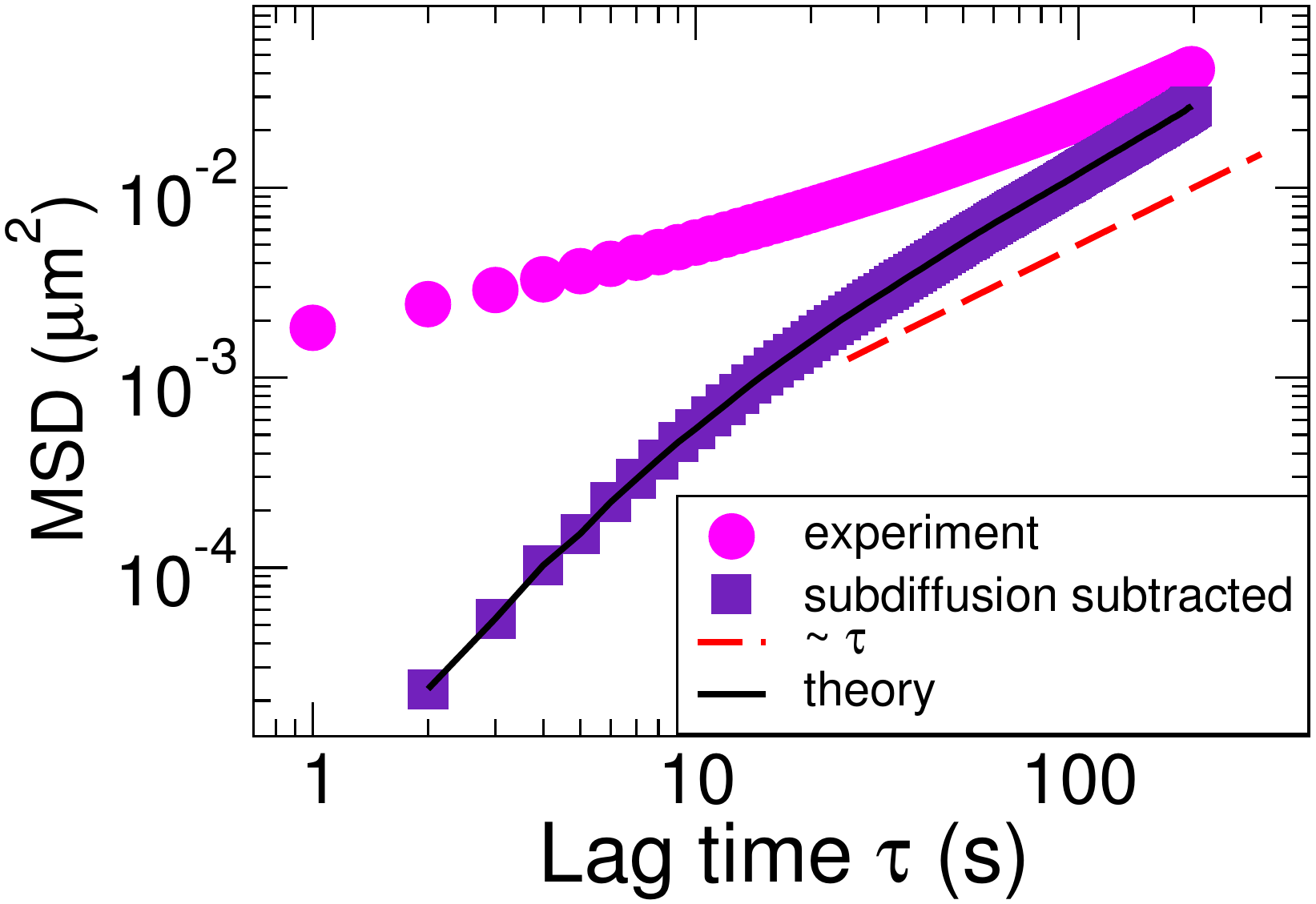}
  \caption{A subtraction procedure based on the model confirms the
    superdiffusive nature of the observed rapid chromosomal
    movements. Purple circles are mean-square displacement vs lag time
    from the same data as in Fig.~\ref{fig:1}. Indigo squares are obtained by
    subtracting the background fBm contribution 
    from the mean-squared
    displacement [Eq.~\eqref{eq:MSDformal}]. The remaining
    contribution to the mean-square displacement is super-linear (the red
    dashed line is the linear scaling), as predicted by Eq.~\eqref{eq:parametri}, consistent
    with the hypothesis of intermittent active relocations.  }
  \label{fig:3}
\end{figure}

As a consistency check of the agreement between model and data, we
considered the possibility of disentangling contributions to the
mean-square displacement from active force and subdiffusion
[Eqs~\eqref{eq:MSDformal} and \eqref{eq:parametri}].

We used this property to define an alternative analysis of the
experimental MSD.  First, we fixed $D_{\textrm{app}}$ and $H$ by
fitting the mean-square displacement for time scales below $10$s
averaged on all tracks.
We then considered the subtracted MSD, calculated as
$\mathrm{MSD}(\tau)-2dD_\mathrm{app}\tau^{2H}$, and fitted it with a
polynomial function (assuming the regime for the lag times $\tau \gg
\tau_0$).  Remarkably, this procedure leads to very similar parameter
values as the blind fit (Table~\ref{tab:params1}B).
By comparison, the best-fitting fractional Brownian motion performs
much worse (Table~\ref{tab:fitresidual}).
The fact that the two procedures lead to essentially the same
parameters suggests that the parameter region that can reproduce the
experimental data is localized and univocal, and the existence of a
null manifold or multiple solutions is unlikely.
In order to further support this statement, 
we performed systematic 4-parameter fits of $\tau_0$, $w$, $V$, and
$D_\mathrm{app}$, by varying $2H$ in the interval $(0,1)$.  Comparing
the goodness-of-fit scores confirmed that there is a single global
best fit, corresponding to $2H=0.404$
(parameters are reported in Table~\ref{tab:params1}C).

\begin{table}
  \centering
\[
\begin{array}{c | c  c  c }	
\ & \textrm{Best-fit fBm} & \textrm{Constr. fit} & \textrm{Joint fit}  \\
\hline
\textrm{Reduced RMS}  & \num{5.9e-03} & \num{2e-06} & \num{7.6e-05} \\
\end{array}
\]
\caption{Comparison of the goodness-of-fit scores for the MSD in the
  data with the best-fitting simple fBm model, with the constrained
  active-force model fit, and with the joint fit keeping into account
  the end-to-end distance distribution
  (Table~\ref{tab:params1}D). The reduced RMS is defined as the
    chi-square residual divided by the number of degrees of freedom.}  
\label{tab:fitresidual}
\end{table}

\subsection*{Model predictions beyond mean-square displacement}

So far we considered constrained fits based exclusively on the MSD
curves fixing \emph{a priori} $H$ or both $D_{\textrm{app}}$ and $H$
based on the short-time behavior, which should estimate the background
process.
This constrained procedure fixes all parameters, and we verified that
the model greatly improves the best fit of MSD vs lag time compared to
a normal fBm (Table~\ref{tab:fitresidual}), by comparing their
  reduced chi-square.
These clear quantitative differences mirror the important
  qualitative difference between the model and the normal fBm. Indeed,
  the latter model simply cannot reproduce the MSD curves observed in
  the data, which are visibly bent in log-log scale.

However, while the ensemble-averaged MSD is useful to establish
  that an improved model is needed, this mean quantity is notoriously
non discriminating.  Importantly, the active-force model also leads to
different predictions for single-track properties, so it could
reproduce the data more effectively than a simple fBm even if its
performance on the MSD fit were equivalent.  Many models, each with a
distinct physical mechanism, may predict the same (bent)
ensemble-averaged MSD.  For example, a ``tempered''
fBm~\cite{Meerschaert2013} (with a time cutoff in the noise
autocorrelation function) would crossover to diffusive behavior at a
prescribed time scale. However, this model would not be able to show
super-diffusive behavior on a subset of tracks, as it is visible in
the data.

Hence, more detailed comparisons with tracking data are useful.  We
considered the single-track end-to-end distance, and we verified that
this gave equivalent result to an effective drift velocity based on
the projection of the end-to-end distance on the main track axis used
in previous work~\cite{Javer2014} (not shown).
%
%
All these observables are independent predictions from the model
(whose parameters are fixed by the fitting procedure of the
ensemble-averaged MSD) and can be matched precisely in track length
and sampling to experimental data. Additionally, these quantities
should discriminate models that predict near-ballistic behavior for a
subset of tracks.

\begin{figure}[t]
  \centering
  \includegraphics[width=0.35\textwidth]{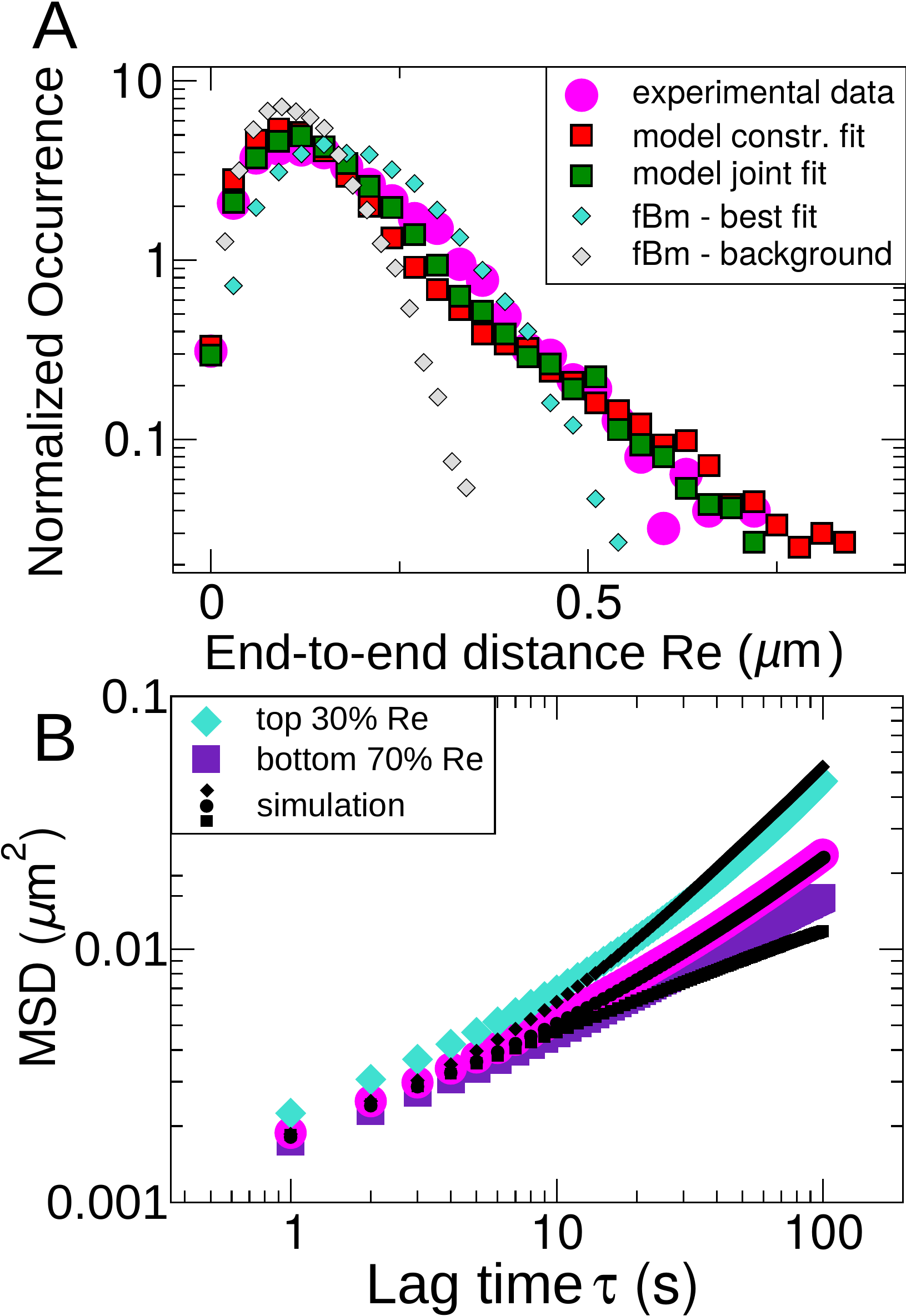}
  \caption{The model predicts correctly the qualitative behavior of
    single-track end-to-end distances and conditional mean-square
    displacements. A: Distribution of the track end-to-end distance
    $R_\mathrm{e}$ in experimental data (purple circles), the
    intermittent active force model (black filled smaller circles) and
    fractional Brownian motion (open smaller circles). Simulations
    have been matched with experimental data, both in number and
    length of tracks (parameters shown in Tab.~\ref{tab:params1}).  B:
    Conditional averages of mean-square displacements on tracks within
    a given percentile of $R_\mathrm{e}$ (diamonds, top $30\%$, and
    squares, bottom $70\%$ shown as example) show qualitative
    agreement and some quantitative discrepancies. Circles are the
    overall MSD average shown in Fig.~\ref{fig:2}.  }
  \label{fig:4}
\end{figure}

Fig.~\ref{fig:4}A shows that the prediction for the distribution of
the track end-to-end distance $R_\mathrm{e}$ improves the estimate of
the tails with respect to the background fBm, as well as with respect
to a best-fit fBm. Fig~\ref{fig:4}B shows mean-square displacements as
a function of lag times, both for all tracks and as conditional
averages on the subset of tracks whose $R_\mathrm{e}$ is in the top
$30\%$ and bottom $70\%$ of the distribution shown in
Fig~\ref{fig:4}A.

The agreement is remarkable, since the model is adjusted only through a
fit of the mean-square displacement, so that the agreement has to be
regarded as an independent prediction.
However, the active force model tends to overestimate the tail of the
end-to-end distance and to underestimate the diffusivity of tracks
where the ballistic transport is not active. The former effect becomes
evident at long lag times, while the latter appears at short lags
(as visible in Fig.~\ref{fig:4}AB).
The best-fitting fBm, while not being able to capture the tails
present in the data, appears to give better a compromise (and is
definitely more parameter-poor) between bulk behavior and
tails. However, this model cannot be considered a viable alternative,
since its performance is clearly much worse in fitting the
ensemble-averaged MSD, and it gives an unrealistic value of the
exponent $2H$, close to 0.8 (Table~\ref{tab:params1}E).

\subsection*{Joint optimization of model parameters on end-to-end
  distance}

To improve the agreement of model with single-track data, we performed
a joint fit where, instead of fixing the parameters purely based on
the analytical MSD fit, we also considered explicitly the long-time
end-to-end distances.  Specifically, we considered all the best fits
of the model at fixed $D_{\mathrm{app}}$ and $H$, for a grid of these
values, and we evaluated the manifold of residuals on the histogram of
end-to end distance $R_\mathrm{e}$. The parameters for the optimal fit with
these criteria are given in Table~\ref{tab:params1}D, and the
resulting distribution of $R_\mathrm{e}$ is plotted in Fig.~\ref{fig:4}A.
Additionally, Table~\ref{tab:fitresidual} shows that the trade-off in
the residuals of the MSD for this fit is acceptable, and gives a
better agreement with the data than the best-fit fBm.

One possible additional source of error overestimating the tails of
the end-to-end track length distribution is the fact that the model
assumes a constant force in direction and orientation for the active
process, which is not plausible in the data. The discrepancy between
model and data is expected to occur when the force is active more than
once in a single track.  With the model parameters, a simple estimate
leads to the expectation for this to happen in 2-4\% of
the tracks (which are 100s long), and thus we can predict that this
factor only affects the tail of the end-to-end distance
distribution in this range of percentiles.

%
%

\subsection*{Different chromosomal loci have coordinate-specific
  parameters }

\begin{figure}[h]
  \centering
  \includegraphics[width=0.35\textwidth]{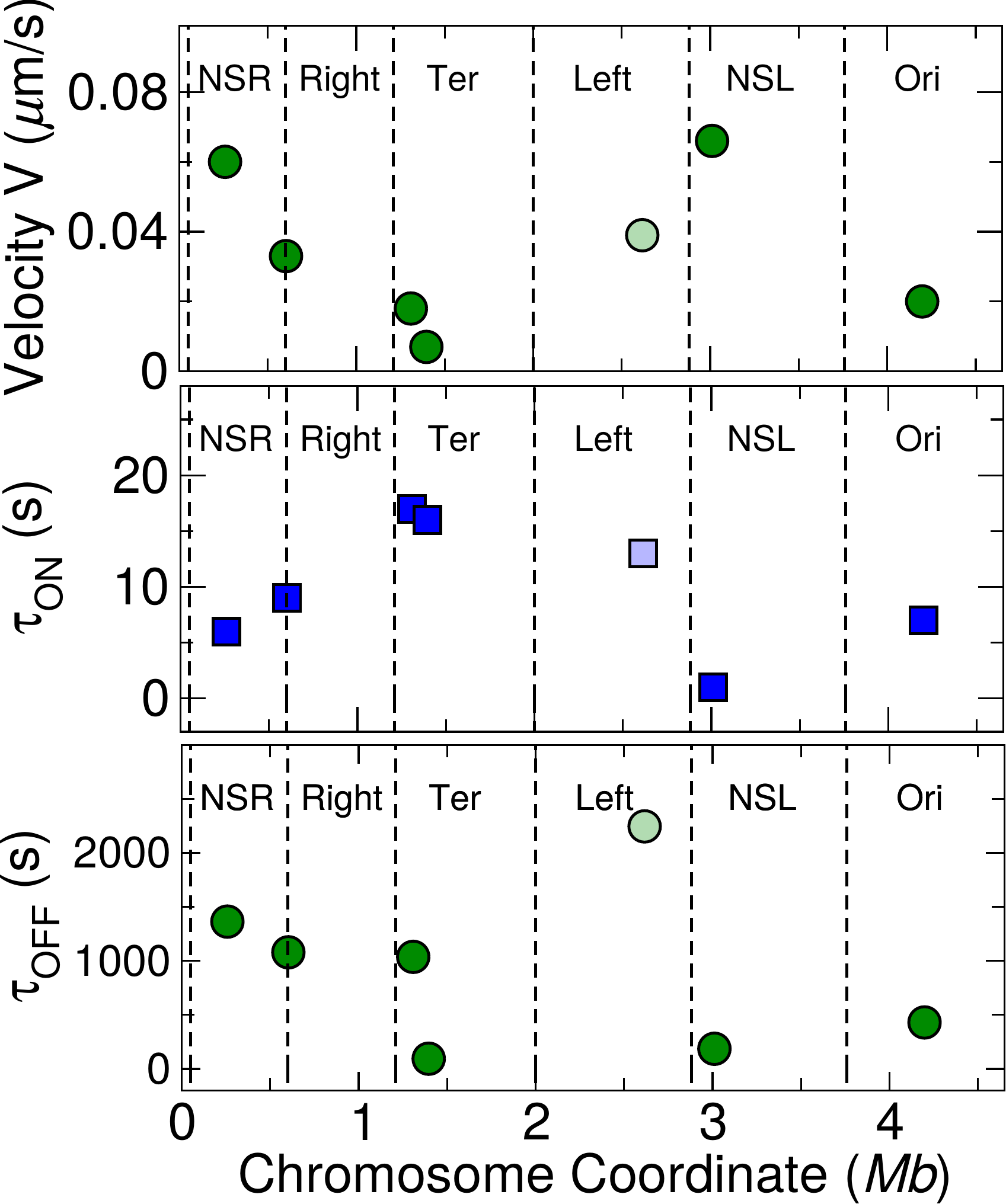}
  \caption{Parameter values for different chromosomal loci. Top panel:
    typical velocity of excursions $V$. Middle panel: characteristic
    time $\tau_\mathrm{on}$. Bottom panel: characteristic time
    $\tau_\mathrm{off}$. The results of the less reliable fit of the
    Left1 locus are shown in faded colors. The strong differences
      observed for different chromosomal loci can be interpreted as
      local differences in the physical organization or in the action
      of the active drive\cite{Javer2013,Javer2014}.  (Estimated
    errors are around the size of the points.)  }
  \label{fig:5}
\end{figure}

The fitting procedures defined above are applicable to data from
different loci. We applied the constrained fit to all loci from the
datasets in refs.\cite{Javer2013,Javer2014} where 1s lag data were
available. The results are shown in Fig.~\ref{fig:5}.  We can obtain a
rough estimate of the errors in these results by considering the range
of variability of the fits analyzed above (Tab.~\ref{tab:params1}),
which is of order $10\%$ for $V$, $\tau_\mathrm{on}$, and
$\tau_\mathrm{off}$: the symbol size in Fig.~\ref{fig:5} reflects the
estimated errors.
The results indicate a
clear trend for the typical velocity $V$, which follows an
opposite trend to $\tau_\mathrm{on}$.
Since the anticorrelation between $V$ and $w$
(Eq.~\eqref{eq:parametri2}) is intrinsic of the model, one may
interpret this trend as the variation of a single physical parameter.
Conversely, the characteristic
off-time $\tau_\mathrm{off}$ does not show a clear trend, and tends to
become very large in some loci.
The differences in the estimated process observed for different
  chromosomal loci can be interpreted as differences in the physical
  organization of the chromosome or in the action of the active
  drive~\cite{Javer2013,Javer2014,Fisher2013}.
In one specific case, the Left1 locus, the algorithm used for the fit
locates a very shallow minimum for $\tau_\mathrm{off}$, so it is
difficult to pinpoint a precise value for this parameter.  We believe
that the applicability of the model to this particular locus is
debatable (the corresponding points are highlighted in
Fig.~\ref{fig:5}. The values for the apparent diffusion constant fits
of the loci (not shown) are coherent with the trends found in
ref.~\cite{Javer2013}.

\subsection*{Active movements carry additional noise}

Finally, we addressed a second shortcoming of the model, visible in
Fig.~\ref{fig:4}.  Namely, the predictions of the conditional
mean-square displacements for tracks with end-to-end distance $R_\mathrm{e}$ in
the higher or lower tails of the distribution show relevant
differences between model and data (Fig.~\ref{fig:4}B). In particular,
the mean-square displacement in the model is essentially independent
of track end-to-end distance $R_\mathrm{e}$ at short lag times, while the data
are not.

The fact that the model should behave this way is evident from
Eq.~(\ref{eq:modello}) and (\ref{eq:parametri2}). At short time lags,
fractional Brownian motion dominates the displacement, and this
process occurs at fixed noise level.
Hence, the model at short lags behaves precisely as an fBm with fixed
noise amplitude, and therefore it cannot show the variation in
diffusivity found in the data. In other words, for short enough lags,
even trajectories where the driving is switched on should show the
same amount of subdiffusion.

\begin{figure}[h]
  \centering
  \includegraphics[width=0.35\textwidth]{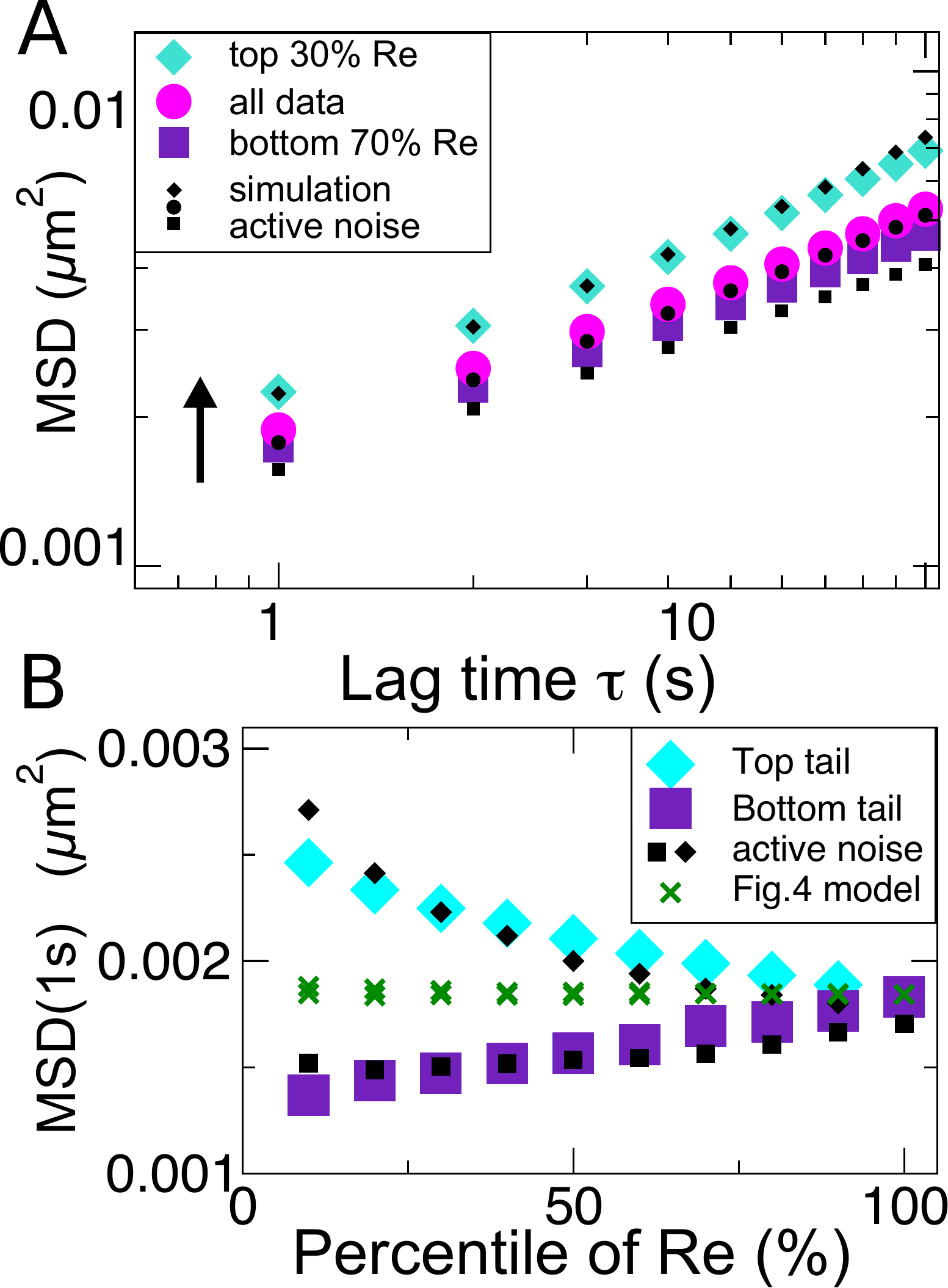}
  \caption{Additional diffusivity of active excursions.  A: the model
    variant including active noise (black filled symbols) fully
    captures the conditional mean-square displacement at short time
    lags of experimental data (large filled symbols), for trajectories
    that fall in different percentiles of end-to-end distance $R_\mathrm{e}$ (top
    30\% $R_\mathrm{e}$, diamonds \emph{versus} bottom 70\%, squares).  B:
    Conditional mean-square displacement at 1s time lag computed as a
    function of end-to-end distance percentile, and compared with
    simulations.  Model parameters in Table~\ref{tab:params1}B, $q=0.78$. 
  }
  \label{fig:6}
\end{figure}

Instead, the lack of agreement between data and model suggests that
active movements may be also characterized by increased noise levels,
on top of a directional driving force.
Physically, this could be due to heterogeneity in the diffusion
coefficient, related to the active excursions (see
below)~\cite{Wang2012}.
In order to explore this hypothesis, we defined a variant of the model,
where active movements are also subject to increased noise
levels. This is described by the following equation of motion
\begin{equation}
  \dot{x_i}(t) = \left[1+ q \sigma(t)\right]\xi_{i,fBm}(t) + V_i\sigma(t) \ ,
  \label{eq:modello_actnoise}
\end{equation}
where $q$ describes a (nonthermal) contribution to noise amplitude in
presence of active motion, and the other quantities are identical to
Eq.~(\ref{eq:modello}).  The extra noise parameter $q$ determines an
increased diffusivity in presence of the driving process and is
  determined by a fit of the extended model.
This modified model makes it qualitatively possible for the MSD to
  vary with track end-to-end distance, as observed in the data.
Fig.~\ref{fig:6}A shows that simulations of this model variant
reproduce the short time-lag changes in the conditional mean-square
displacement for trajectories with varying end-to-end distance.
To better capture the specificity of this model variant,
Fig.~\ref{fig:6}B shows the conditional mean-square displacement at 1s
time lag for tracks in the bottom and top tail of the end-to-end
distance distribution, plotted as a function of end-to-end distance
percentile.  In this plot, the value of the $x$ axis refers to the
$x-\%$ top and bottom tail of the track end-to-end distance
distribution. Hence, the plot tests the consistency of the agreement
between model and data if the cutoff is shifted lower or higher in the
end-to-end distance distribution.
Comparison of data with simulations of both models shows that only the
variant with modified noise during active stretches is able to
modulate the diffusivity at short lags.
%
%
%

\section{Discussion and Conclusions. }
\label{sec:Disc}

We considered a phenomenological model of
subdiffusing particles performing fractional Brownian motion and
subject to intermittent ballistic excursions and analyzed  the
validity of this description for chromosomal loci movements.
The comparison procedure allows to evaluate the relevant time scales
and to dissect some specific features of the data.
The model defined here potentially has a wider range of applications,
including eukaryotic chromosomal loci, where similar active phenomena
have been reported, and may or may not have the same
interpretation~\cite{Bronshtein2016,Bronstein2009}.

The first outcome of the model concerns the estimated intensity and
typical time scales of the active movements. Different approaches and
fits all suggest that the characteristic times of active relocations
span a few seconds, while the typical waiting times between
activations are of the order of a few hundred seconds. Both processes
occur below the typical time scales of a cell interdivision time (tens
of minutes to hours) and hence should be observable in every cell, and
overlap with the key cell processes of replication and chromosome
segregation.  Additionally, the typical speeds of the relocations are
estimated to be slightly over one micron per minute. These figures are
in agreement with previous
reports~\cite{Kleckner2014,Javer2014,Fisher2013,joshi11}, but the
present work is a systematic attempt to capture these time scales
quantitatively using a theoretical framework.

A second important feature of the model is its ability to generate
nontrivial testable predictions.  We first used the mean-square
displacement to fix the parameters, and then considered the behavior
of the distributions of track end-to-end distances
and the conditional mean-square displacements.
The model captures the behavior of these quantities in experimental
data better than the best-fit fBm. However, some discrepancies exist
both at short and at long time scales.
A model fit purely based on the mean-square
displacement behavior tends to overestimate the mobility of the tracks
where the transport is switched on, and this effect is particularly
visible at long time scales. To compensate for this behavior, we
defined a fitting procedure that also takes into account the
end-to-end distance distributions at these long time scales, which
gives a more satisfactory agreement.
Additional differences at long time scales may be attributed to
simplifying hypotheses in the definition of the model, such as the
assumption of a single relevant scale for the diving force and of a
simple telegraph process for the force switch.
Focusing on short time scales, we have isolated an increased
diffusivity of faster-moving foci as a potential relevant ingredient
(see below).

Notably, we have compared different models with different number
  of parameters. Admittedly, it may seem unsurprising that models with
  more parameters give better results. The important feature to note
  is that each extension considered here is defined based on the
  ability to capture a different qualitative behavior. For example,
  the intermittent force model can produce MSD curves that are bent in
  a log-log scale, which is impossible with a normal two-parameter
  fBm.  Alternative models that may include this qualitative feature
  (such as a tempered fBm) would still entail adding more
  parameters. 
  Equally, the model variant with additional noise in active movements
  can reproduce the qualitative feature of increased diffusivity in
  tracks with increasing end-to-end distance, which is not possible to
  reproduce in the simpler variant of the model.

Our analysis of different fitting procedures (compared in Table
\ref{tab:params1}) can be used to produce a rough estimate of the
errors on these parameters. Considering the values of the different
fits, we expect the errors to be between 5\% and 10\% for the typical
velocity $V$, and the transition time scales $\tau_{\mathrm{on}}$ and
$\tau_{\mathrm{off}}$, and less than $1\%$ for the scaling exponent
$H$ and the apparent diffusion constant $D_{\mathrm{app}}$.

It is interesting to compare the value of the inferred model
parameters with some rates and durations of relevant biological
processes at play.
The characteristic times for active relocations (5--10 s) agree well
with the time scales of the pulses of density shift~\cite{Fisher2013}
observed along the nucleoid ($\sim$5 s). These pulses were found to
occur at about 20 minutes intervals, to be compared with the 7-23
minutes of the fitted characteristic off times. From our fits, these
values appear to be locus dependent, and to be closer to 20 minutes in
the Right-Ter arm of the chromosome. Finally, the characteristic speed
of the active process, about a micron per minute, compares well with
the speed of these fast processes~\cite{Javer2014,joshi11,Fisher2013},
happening at much faster time scales than the average speed of
segregation, which is on the scale of microns per hour.
These characteristic times vary along the chromosome, coherently with
previous findings~\cite{Javer2014,Javer2013} that suggested
differential organization and/or local noise along the chromosomal
coordinates. For some loci, the characteristic off-time of the process
becomes very large, indicating that the active excursions could become
extremely rare.

Finally - the comparison of model predictions and data leads us to the
conclusion that active relocation also carries increased noise levels.
Specifically, a model where noise amplitude is constant in presence of
active excursions cannot reproduce the increase in diffusivity at
short lag times observed in experimental data. This suggests that the
processes generating the active relocations are concurrent with the
noise-increasing processes.
The microscopic interpretation of this result is unclear. One
possibility is that this increased diffusivity is due to 
nonthermal active fluctuations~\cite{theriot12}.
However, nonthermal random forces are expected to dominate at long
time lags.
The noise increase has been proven to be ATP-dependent, but
not associated to any specific process such as the activity to DNA
gyrase (Topoisomerase favoring relaxation of positive supercoiling) or
depolymerization of MreB (cell wall biosynthesis), and is only weakly
linked to RNA polymerase activity. 
Active relocations have been previously interpreted as relaxation of stress
(generated by process such as DNA replication and transcription) due
to release of internal tethering interactions (e.g., by bridging
proteins such as H-NS and Fis or condensins such as
MukBEF)~\cite{Kleckner2014,Javer2014,Fisher2013}. This kind of motion
is not necessarily associated to increased noise, because the
stress-release events might be well-separated temporally from the
stress-generating ones.

Another possibile interpretation (complementary to the previous one)
of the increased mobility in presence of active movements might
explain this behavior.  This is related to the reported glassy
properties of the cytoplasm system~\cite{Parry2014}, which should also
affect the nucleoid. In this framework, active relocation might cause
a fluidization effect, releasing local portions of cytoplasm and
nucleoid from ``cages'' where they are otherwise confined with limited
mobility. In this case, the differential noise would be due to
heterogeneity in the sub-diffusion process~\cite{Wang2012,Wang2009}.
Possibly more general sources of heterogeneity than glassyness could
also lead to similar effects.
Such kind of disorder has been recently implicated for the motion of
cytoplasmic particles~\cite{Lampo2017}.
A more precise dissection of such hypothesis requires a theoretical
approach that incorporates explicitly the more complex physical
ingredient of crowding and glassy behavior.

\section{Acknowledgments} 
We are grateful to Pietro Cicuta, Kevin Dorfman, Alessandro Taloni for
helpful discussions.
This work was supported by the International Human Frontier Science
Program Organization, grant RGY0070/2014. MT is grateful for the
financial support of the EU-Horizon2020 IRSES program DIONICOS
(612707).

\bibliography{Rapid_telegraph.bib} 

\end{document}